\documentclass[aps,twocolumn,superscriptaddress,tightenlines]{revtex4}
\usepackage{epsfig,graphicx,times}
\usepackage{amstext}
\usepackage{amsmath}
\usepackage{amssymb}
\usepackage{graphicx}
\usepackage{latexsym}
\usepackage{bm}
\usepackage[colorlinks,citecolor=blue, linkcolor=blue,hyperindex,CJKbookmarks,dvipdfm]{hyperref}

\begin{document}

\title{Single-photon nonreciprocal transport in one-dimensional coupled-resonator waveguides}
\author{Xun-Wei Xu}
\email{davidxu0816@163.com}
\affiliation{Department of Applied Physics, East China Jiaotong University, Nanchang, 330013, China}
\author{Ai-Xi Chen}
\email{aixichen@ecjtu.edu.cn}
\affiliation{Department of Physics, Zhejiang Sci-Tech University, Hangzhou, 310018, China}
\affiliation{Department of Applied Physics, East China Jiaotong University, Nanchang, 330013, China}
\author{Yong Li}
\affiliation{Beijing Computational Science Research Center, Beijing 100193, China}
\affiliation{Synergetic Innovation Center of Quantum Information and Quantum Physics,
University of Science and Technology of China, Hefei 230026, China}
\affiliation{Synergetic Innovation Center for Quantum Effects and Applications, Hunan
Normal University, Changsha 410081, China}
\author{Yu-xi Liu}
\affiliation{Institute of Microelectronics, Tsinghua University, Beijing 100084, China}
\affiliation{Tsinghua National Laboratory for Information Science and Technology
(TNList), Beijing 100084, China}
\date{\today }

\begin{abstract}
We study the transport of a single photon in two coupled
one-dimensional semi-infinite coupled-resonator waveguides (CRWs), in which
both end sides are coupled to a dissipative cavity. We demonstrate that a
single photon can transfer from one semi-infinite CRW to the other
nonreciprocally. Based on such nonreciprocity, we further construct a three-port single-photon circulator by a T-shaped waveguide,
in which three semi-infinite CRWs are pairwise mutually coupled to each other. The single-photon nonreciprocal transport is
induced by the breaking of the time-reversal symmetry and the optimal
conditions for these phenomena are obtained analytically. The CRWs with
broken time-reversal symmetry will open up a kind of quantum devices with
versatile applications in quantum networks.
\end{abstract}

\maketitle

\section{Introduction}

The realization of quantum networks, which are composed of many quantum nodes and quantum channels,
is one of the main goals in quantum
information science~\cite{KimbleNat08a}. Quantum nodes are used to
generate, process and store quantum information. Quantum channels, connected by quantum nodes, are used to transmit quantum states across the entire network. The coupled-resonator waveguide (CRW) with
good scalability and integrability~\cite%
{WallraffNat04,NotomiNPT08,CastellanosAPL07,AbdumalikovPRL10a} provides an
appropriate platform for studying quantum state transmission in quantum
networks. Currently, single-photon transports in one-dimensional CRWs
coupled to different kinds of quantum nodes are studied theoretically~%
\cite%
{LZhouPRL08a,LZhouPRA08a,JQLiaoPRA09a,LZhouPRA09a,LongoPRL10a,JQLiaoPRA10a,ZHWangPRA12a,LZhouPRA12a,MTChengPRa12a,ZRGongPRA08a,LZhouPRL13a,JLuPRA14a,JLuOE15a,ZHWangPRA14a}%
, and many important quantum devices are proposed, such as quantum switch~%
\cite%
{LZhouPRL08a,LZhouPRA08a,JQLiaoPRA09a,LongoPRL10a,LZhouPRA09a,JQLiaoPRA10a,ZHWangPRA12a,LZhouPRA12a,MTChengPRa12a}%
, photon memory~\cite{ZRGongPRA08a}, single-photon router~\cite%
{LZhouPRL13a,JLuPRA14a,JLuOE15a}, and frequency converter~\cite{ZHWangPRA14a}%
. In all these studies, the quantum devices based on CRWs are reciprocal. However, we know that
unidirectional devices such as isolators and circulators are indispensable
elements to realize quantum networks.
It has already been shown that the breaking of Lorentz reciprocity is pivotal for isolators~\cite{JalasNPT13}.
One class of nonreciprocal systems which can be used for isolators and circulators is based on the breaking of the time-reversal symmetry.

In general, the breaking of time-reversal symmetry in optical systems can be
generated by two different ways: (i) using magneto-optical effects (e.g.,
Faraday rotation)~\cite%
{FujitaAPL00,EspinolaOL04,ZamanAPL07,HaldanePRL08,ShojiAPL08,ZWangNat09,HadadPRL10,KhanikaevPRL10,LBiNPo11,ShojiOE12,ZWangSR15a,YShiNPt15a,XGuoPRL16,QTCaoPRL17,DelBinoarX16a}
and (ii) non-magnetic strategies (e.g., employing optical nonlinearity~\cite%
{GalloAPL01,MingaleevJOSAB02,SoljacicOL03,RostamiOLT07,AlberucciOL08,LFanSci12,LFanOL13,AnandNL13,BiancalanaJAP08,MiroshnichenkoAPL10,CWangOE11,CWangSR12,KXiaOE13,LenferinkOE14,YYuarX14}, dynamical modulation~\cite%
{YuNP09,KFangNPo12,ELiNC14,DoerrOL11,DoerrOE14,LiraPRL12,KFangPRL12,ManipatruniPRL09,TzuangNPt14,MunozPRL14,YYangOE14,WangOE10,MSKangNP11,EuterNP10,RamezaniPRA10,LFengSci11,BPengNP14,WangPRL13,JHWuPRL14,HorsleyPRL13,HafeziOE12,KimNPy15,CHDongNC15,EstepNPy14a,ZShenNPt16a}, etc). Especially, as a non-magnetic strategy, optical nonreciprocity in the
coupled cavity modes with relative phase has drawn more and more attentions
in recent years, and many different structures have been proposed
theoretically~\cite%
{KochPRA10,HabrakenNJP12,RanzaniNJP14a,RanzaniNJP15a,YPWangSR15a,XuXWPRA15,SliwaPRX15,SchmidtOpt15,MetelmannPRX15,FangKArx15,XWXuPRA16a,MiriarX16a,MetelmannarX16a,LTianarx16a,XXuarX17a}
or demonstrated experimentally~\cite{RuesinkNC16a,KFangNPy17a,BernierarX16a}.
The isolators and circulators, made of three or four coupled cavity modes,
can be viewed as a quantum node in the quantum network.
However, the input and output fields in these nonreciprocal devices are usually treated by the input-output relations~\cite{Gardiner}.
The detailed construction of the quantum channels
coupled to the isolators and circulators are not considered seriously.

In this paper, we study single-photon nonreciprocal transport in CRWs.
In Sec.~II, we study the transport of a single photon in a waveguide
consisting of two coupled semi-infinite CRWs, in which both end sides are coupled to a dissipative
cavity, and show that a single photon can transfer from one
semi-infinite CRW to the other nonreciprocally due to the breaking of the
time-reversal symmetry. In Sec.~III, a T-shaped waveguide is proposed by
replacing the dissipative cavity discussed in Sec.~II with another semi-infinite CRW, and we
further demonstrate that the T-shaped waveguide can be used as a three-port
single-photon circulator in the broken regime of the time-reversal symmetry.
Finally, the main results of the paper are summarized in Sec.~VI.

\section{Single-photon nonreciprocity in two semi-infinite CRWs}

\subsection{Theoretical model and scattering matrix}

\begin{figure}[tbp]
\includegraphics[bb=0 452 589 645, width=8.5 cm, clip]{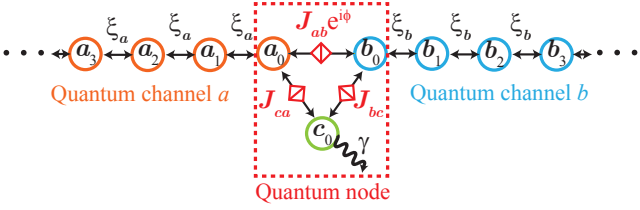}
\caption{(Color online) Schematic diagram of a one-dimensional waveguide consisting of two coupled semi-infinite
CRWs ($a_{j}$ and $b_{j}$ for $j\geq 0$), in which both end sides ($a_{0}$ and $b_{0}$) are coupled to a dissipative cavity
($c_{0}$).}
\label{fig1}
\end{figure}

As schematically shown in Fig.~\ref{fig1}, a one-dimensional waveguide consists of two coupled semi-infinite
CRWs, in which both end sides are coupled to a dissipative cavity.
Each semi-infinite CRW is made of single-mode
cavities, coupled to each other through coherent hopping of photons between
neighbouring cavities. The CRWs are quantum channels for single-photon
transmission and the three coupled cavities denoted by $a_{0}$, $b_{0}$ and $%
c_{0}$ are served as a quantum node for single-photon scattering.
We assume that the two semi-infinite CRWs have no dissipation.
The Hamiltonian of the system can be described by%
\begin{equation}\label{Eq1}
H=\sum_{l=a,b,c}H_{l}+H_{\mathrm{int}}.
\end{equation}%
The Hamiltonians for the two CRWs (quantum channels) are
\begin{equation}
H_{a}=\omega _{a}\sum_{j=0}^{+\infty }a_{j}^{\dag }a_{j}-\xi
_{a}\sum_{j=0}^{+\infty }\left( a_{j}^{\dag }a_{j+1}+\mathrm{H.c.}\right),
\end{equation}%
\begin{equation}
H_{b}=\omega _{b}\sum_{j=0}^{+\infty }b_{j}^{\dag }b_{j}-\xi
_{b}\sum_{j=0}^{+\infty }\left( b_{j}^{\dag }b_{j+1}+\mathrm{H.c.}\right),
\end{equation}%
where $l_{j}$ ($l_{j}^{\dag }$, $l=a,b$) is the bosonic annihilation
(creation) operator of the $j$th cavity in the CRW-$l$ with the same
resonance frequency $\omega _{l}$, and the coupling strength between two
nearest-neighbor cavities $\xi _{l}$ is also the same in the CRW-$l$.
The Hamiltonian for the dissipative cavity is
\begin{equation}\label{Eq4}
H_{c}=\left( \omega _{c}-i\gamma \right) c_{0}^{\dag }c_{0},
\end{equation}%
where $c_{0}$ ($c_{0}^{\dag }$) is the bosonic annihilation (creation) operator of the
dissipative cavity with resonance frequency $\omega _{c}$ and damping rate $%
\gamma $.
The interaction terms in the quantum node is
\begin{eqnarray}\label{Eq5}
H_{\mathrm{int}} &=&J_{ab}(e^{i\phi }a_{0}^{\dag }b_{0}+e^{-i\phi
}a_{0}b_{0}^{\dag })  \notag \\
&&+J_{bc}(b_{0}^{\dag }c_{0}+b_{0}c_{0}^{\dag })  \notag \\
&&+J_{ca}(c_{0}^{\dag }a_{0}+c_{0}a_{0}^{\dag }),
\end{eqnarray}%
where $J_{ab}e^{i\phi _{ab}}$, $J_{bc}e^{i\phi _{bc}}$ and $J_{ca}e^{i\phi _{ca}}$ are the coupling
constants between the cavities $a_{0}$, $b_{0}$ and $c_{0}$ with real strengths ($J_{ab}$, $J_{bc}$ and $J_{ca}$) and phases
($\phi_{ab}$, $\phi_{bc}$ and $\phi_{ca}$), and only the total phase $\phi =\phi _{ab}+\phi
_{bc}+\phi _{ca}$ has physical effects. Without loss of
generality, $\phi $ is only kept in the terms of $a_{0}^{\dag }b_{0}$ and $%
a_{0}b_{0}^{\dag }$ in Eq.~(\ref{Eq5}) and the following derivation. It should be
noted that the time-reversal symmetry of the whole system (without dissipation, i.e., $\gamma=0$) is broken when we
choose the phase $\phi\neq n\pi$ ($n$ is an integer). As we will show in the
following, $\phi\neq n\pi$ is one of the crucial conditions to demonstrate
single-photon nonreciprocity.

The stationary eigenstate of single-photon scattering in the waveguide is
given by
\begin{equation}\label{Eq6}
\left\vert E\right\rangle =\sum_{l=a,b}\sum_{j=0}^{+\infty }u_{l}\left(
j\right) l_{j}^{\dag }\left\vert 0\right\rangle +u_{c}\left( 0\right)
c_{0}^{\dag }\left\vert 0\right\rangle
\end{equation}%
where $\left\vert 0\right\rangle $ indicates the vacuum state of the whole
system, $u_{l}\left( j\right) $ denotes the probability amplitude in the
state with a single photon in the $j$th cavity of the CRW-$l$ and $%
u_{c}\left( 0\right) $ denotes the probability amplitude with a
single photon in the dissipative cavity. The dispersion relation of the
semi-infinite CRW-$l$ is given by~\cite{LZhouPRL13a}
\begin{equation}
E_{l}=\omega _{l}-2\xi _{l}\cos k_{l}, \quad 0<k_{l}<\pi,
\end{equation}%
where $E_{l}$ and $k_{l}$ are the energy and wave number
of the single photon in the CRW-$l$ with the bandwidth $4\xi _{l}$. Without loss of
generality, we assume $\xi _{l}>0$ in this paper. It should be pointed out that when the energy $E$ of the incident single photon is out of the energy band of the CRW-$l$, i.e. $E<\omega _{l}-2\xi _{l}$ or $E>\omega _{l}+2\xi _{l}$, then the wave number $k_{l}$ becomes a complex number and this single photon can not transport freely in the CRW-$l$.

Substituting the stationary eigenstate
in Eq.~(\ref{Eq6}) and the Hamiltonian in Eq.~(\ref{Eq1}) into the eigenequation $H\left\vert
E\right\rangle =E\left\vert E\right\rangle $, we can obtain the coupled
equations for the probability amplitudes in the quantum node as%
\begin{equation}\label{Eq8}
(\omega _{a}-E)u_{a}\left( 0\right) -\xi _{a}u_{a}\left( 1\right)
+J_{ab}e^{i\phi }u_{b}\left( 0\right) +J_{ca}u_{c}\left( 0\right) =0 ,
\end{equation}
\begin{equation}\label{Eq9}
(\omega _{b}-E)u_{b}\left( 0\right) -\xi _{b}u_{b}\left( 1\right)
+J_{ab}e^{-i\phi }u_{a}\left( 0\right) +J_{bc}u_{c}\left( 0\right) =0 ,
\end{equation}
\begin{equation}\label{Eq10}
\left( \omega _{c}-E-i\gamma \right) u_{c}\left( 0\right) +J_{ca}u_{a}\left(
0\right) +J_{bc}u_{b}\left( 0\right) =0,
\end{equation}
and the coupled equations for the probability amplitudes in the quantum
channels as
\begin{equation}\label{Eq11}
\omega _{l}u_{l}\left( j\right) -\xi _{l}u_{l}\left( j+1\right) -\xi
_{l}u_{l}\left( j-1\right) =Eu_{l}\left( j\right)
\end{equation}%
with $j>0$ and $l=a,b$.

If a single photon with energy $E$ is incident from the infinity side of CRW-%
$l$, the interactions among cavities $a_{0}$, $b_{0}$ and $c_{0}$ in the quantum node will result in photon sacttering
between different CRWs or absorbed by the dissipative cavity. The general
expressions of the probability amplitudes in the CRWs ($j\geq 0$) are given
by
\begin{equation}\label{Eq12}
u_{l}\left( j\right) =e^{-ik_{l}j}+s_{ll}e^{ik_{l}j},
\end{equation}
\begin{equation}\label{Eq13}
u_{l^{\prime }}\left( j\right) =s_{l^{\prime }l}e^{ik_{l^{\prime }}j},
\end{equation}
where $s_{l^{\prime }l}$ denotes the scattering amplitude from the CRW-$l$
to the CRW-$l^{\prime }$ ($l,l^{\prime }=a,b$). Substituting Eqs.~(\ref{Eq12}) and (\ref{Eq13}) into Eqs.~(\ref{Eq8})-(\ref{Eq11}) then we obtain the scattering matrix as%
\begin{equation}
S=\left(
\begin{array}{cc}
s_{aa} & s_{ab} \\
s_{ba} & s_{bb}%
\end{array}%
\right),
\end{equation}%
where
\begin{eqnarray}
s_{aa} &=&D^{-1}\left[ \left( J_{ab}e^{-i\phi }+J_{ba,\mathrm{eff}}\right)
\left( J_{ab}e^{i\phi }+J_{ba,\mathrm{eff}}\right) \right.  \notag \\
&&\left. -\left( \xi _{a}e^{ik_{a}}+J_{aa,\mathrm{eff}}\right) \left( \xi
_{b}e^{-ik_{b}}+J_{bb,\mathrm{eff}}\right) \right],
\end{eqnarray}%
\begin{equation}\label{Eq16}
s_{ab}=D^{-1}\left( \xi _{b}e^{ik_{b}}-\xi _{b}e^{-ik_{b}}\right) \left(
J_{ab}e^{i\phi }+J_{ba,\mathrm{eff}}\right),
\end{equation}%
\begin{equation}\label{Eq17}
s_{ba}=D^{-1}\left( \xi _{a}e^{ik_{a}}-\xi _{a}e^{-ik_{a}}\right) \left(
J_{ab}e^{-i\phi }+J_{ba,\mathrm{eff}}\right),
\end{equation}%
\begin{eqnarray}
s_{bb} &=&D^{-1}\left[ \left( J_{ab}e^{-i\phi }+J_{ba,\mathrm{eff}}\right)
\left( J_{ab}e^{i\phi }+J_{ba,\mathrm{eff}}\right) \right.  \notag \\
&&\left. -\left( \xi _{a}e^{-ik_{a}}+J_{aa,\mathrm{eff}}\right) \left( \xi
_{b}e^{ik_{b}}+J_{bb,\mathrm{eff}}\right) \right],
\end{eqnarray}%
\begin{eqnarray}
D &=&\left( \xi _{a}e^{-ik_{a}}+J_{aa,\mathrm{eff}}\right) \left( \xi
_{b}e^{-ik_{b}}+J_{bb,\mathrm{eff}}\right)  \notag \\
&&-\left( J_{ab}e^{-i\phi }+J_{ba,\mathrm{eff}}\right) \left( J_{ab}e^{i\phi
}+J_{ba,\mathrm{eff}}\right),
\end{eqnarray}%
with the effective coupling strengths $J_{ll^{\prime},\mathrm{eff}}$ induced by the
dissipative cavity defined by
\begin{eqnarray}
J_{ba,\mathrm{eff}} &=&\frac{J_{bc}J_{ca}}{E-\omega _{c}+i\gamma }, \\
J_{aa,\mathrm{eff}} &=&\frac{J_{ca}^{2}}{E-\omega _{c}+i\gamma }, \\
J_{bb,\mathrm{eff}} &=&\frac{J_{bc}^{2}}{E-\omega _{c}+i\gamma }.
\end{eqnarray}%
By setting the incident flow as unit, we define the scattering flows of the
single photons from CRW-$l$ to the CRW-$l^{\prime }$ as the square modulus
of the scattering amplitudes $s_{l^{\prime }l}$ multiplying the group
velocity rates in the corresponding CRWs as~\cite{ZHWangPRA14a}
\begin{equation}
I_{l^{\prime }l}=|s_{l^{\prime }l}|^{2}\frac{\xi _{l^{\prime }}\sin
k_{l^{\prime }}}{\xi _{l}\sin k_{l}}.
\end{equation}
In our model, $I_{ba}\neq I_{ab}$ represents the appearance of nonreciprocal
response, and the perfect nonreciprocity is obtained when $I_{ba}=1$ and $%
I_{ab}=0$, or $I_{ba}=0$ and $I_{ab}=1$.

\subsection{Single-photon nonreciprocity}

Let us now study the optimal conditions for
single-photon nonreciprocity analytically. For simplicity, we assume that
the two CRWs have the same parameters (i.e. $\omega _{a}=\omega _{b}$, $\xi
\equiv \xi _{a}=\xi _{b}$, and $k\equiv k_{a}=k_{b}$) and they are
symmetrically coupled to the dissipative cavity ($J_{c}\equiv J_{ca}=J_{bc}$%
). From Eqs.~(\ref{Eq16}) and (\ref{Eq17}), we can see that $s_{ab}\neq s_{ba}$ (i.e., $%
I_{ab}\neq I_{ba}$) in the case $\phi \neq n\pi $ ($n$ is an integer). The
conditions for perfect nonreciprocity ($I_{ba}=1$ and $I_{ab}=0$, or $%
I_{ba}=0$ and $I_{ab}=1$) are obtained from Eqs.~(\ref{Eq16}) and (\ref{Eq17}) as
\begin{equation}
J_{ab}=\xi,
\end{equation}%
\begin{equation}\label{Eq25}
k=\left\{
\begin{array}{cc}
\phi, & 0<\phi <\pi, \\
2\pi -\phi, & \pi <\phi <2\pi,%
\end{array}%
\right.
\end{equation}%
\begin{equation}\label{Eq26}
\phi =\sin ^{-1}\left( \frac{\gamma \xi }{J_{c}^{2}}\right) \quad \mathrm{{or%
}} \quad 2\pi -\sin ^{-1}\left( \frac{\gamma \xi }{J_{c}^{2}}\right),
\end{equation}%
\begin{equation}\label{Eq27}
\Delta =\xi ^{-1}\left( J_{c}^{2}-2\xi ^{2}\right) \cos \phi,
\end{equation}%
where $\Delta \equiv \omega _{c}-\omega _{a}=\omega _{c}-\omega _{b}$ is the
frequency detuning between the dissipative cavity and the cavities in the
CRWs. To ensure that $\phi $ from Eq.~(\ref{Eq26}) is real, we should choose the parameters $\gamma \xi
\leq J_{c}^{2}$.

\begin{figure}[tbp]
\includegraphics[bb=125 158 465 580, width=8.5 cm, clip]{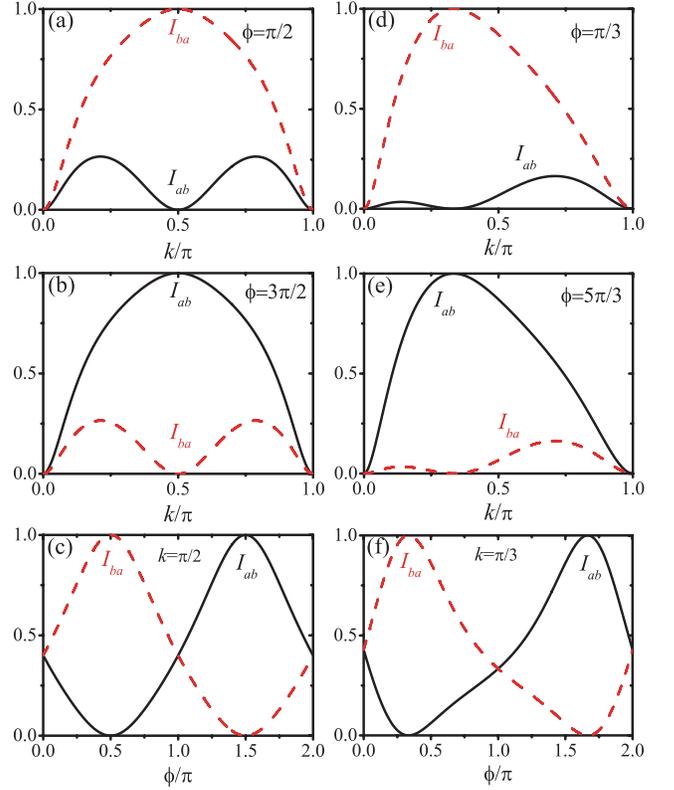}
\caption{(Color online) Scattering flows $I_{ab}$ (black solid curves) and $%
I_{ba}$ (red dashed curves) as functions of the wave number $k/\protect\pi$
for different phases: (a) $\protect\phi=\protect\pi/2$, (b) $\protect\phi=3%
\protect\pi/2$, (d) $\protect\phi=\protect\pi/3$ and (e) $\protect\phi=5%
\protect\pi/3$. Scattering flows $I_{ab}$ (black solid curves) and $I_{ba}$
(red dashed curves) as functions of the phase $\protect\phi/\protect\pi$ for
(c) $k=\protect\pi/2$ and (f) $k=\protect\pi/3$. $J_{c}=\protect\gamma=%
\protect\xi$ for (a)-(c); $J_{c}=\protect\sqrt{2}\protect\xi$ and $\protect%
\gamma=\protect\sqrt{3}\protect\xi$ for (d)-(f). The other parameters are $%
\Delta=0$ and $J_{ab}=\protect\xi$.}
\label{fig2}
\end{figure}

Scattering flows $I_{ab}$ (black solid curve) and $I_{ba}$ (red dashed
curve) as functions of the wave number $k/\pi$ are shown in Figs.~\ref{fig2}%
(a), \ref{fig2}(b), \ref{fig2}(d), and \ref{fig2}(e). The perfect
nonreciprocity appears at the point $k=\phi$ ($0<\phi<\pi$) or $k=2\pi-\phi$
($\pi<\phi<2\pi$), which exhibits good agreement with the analytical result
shown in Eq.~(\ref{Eq25}). Scattering flows $I_{ab}$ (black solid curve) and $I_{ba}$
(red dashed curve) as functions of the phase $\phi/\pi$ are shown in Figs.~%
\ref{fig2}(c) and \ref{fig2}(f), which demonstrate $I_{ba}>I_{ab}$ for $%
0<\phi<\pi$ and $I_{ba}<I_{ab}$ for $\pi<\phi<2\pi$.

\begin{figure}[tbp]
\includegraphics[bb=16 24 562 804, width=8.5 cm, clip]{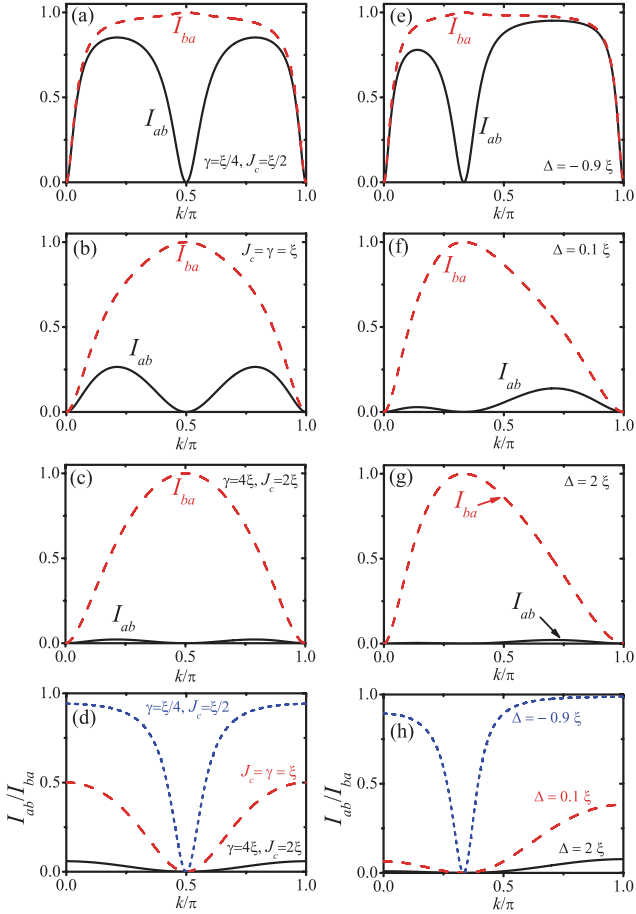}
\caption{(Color online) Scattering flows $I_{ab}$ (black solid curves) and $%
I_{ba}$ (red dashed curves) as functions of the wave number $k/\protect\pi$
in (a)-(c) for $\Delta=0$ and $\protect\phi=\protect\pi/2$, and in (e)-(g)
for $\Delta \neq 0$ and $\protect\phi=\protect\pi/3$. (a) $\protect\gamma=%
\protect\xi/4$ and $J_{c}=\protect\xi/2$; (b) $\protect\gamma=\protect\xi $
and $J_{c}=\protect\xi$; (c) $\protect\gamma=4\protect\xi$ and $J_{c}=2%
\protect\xi$. $\protect\gamma$ and $J_{c}$ are given in Eqs.~(\ref{Eq29}) and (\ref{Eq30}) with $%
\protect\phi=\protect\pi/3$, for $\Delta=-0.9\protect\xi$ in (e), $\Delta=0.1%
\protect\xi$ in (f) and $\Delta=2\protect\xi$ in (g). The ratio of
scattering flow $I_{ab}/I_{ba}$ as a function of the wave number $k/\protect%
\pi$ in (d) for $\Delta=0$ and $\protect\phi=\protect\pi/2$ with the
corresponding parameters $\protect\gamma$ and $J_{c}$ given in (a)-(c), and
in (h) for $\Delta \neq 0$ and $\protect\phi=\protect\pi/3$ with the
corresponding parameters given in (e)-(g). The parameter $J_{ab}=\protect\xi$%
.}
\label{fig3}
\end{figure}

From Eq.~(\ref{Eq26}), when $\phi=\pi/2$ or $3\pi/2$, we can derive that $J_{c}$,
$\gamma$, and $\xi$ satisfy the condition
\begin{equation}\label{Eq28}
J_{c}^{2}=\gamma \xi.
\end{equation}
In Figs.~\ref{fig3}(a)-(c), we show the scattering flows $I_{ab}$ (black solid
curves) and $I_{ba}$ (red dashed curves) as functions of the wave number $%
k/\pi $ when $J_{c}$ and $\gamma$ are taken three different sets of
parameters under the condition in Eq.~(\ref{Eq28}). The scattering flows $I_{ab}$ (black solid
curves) and $I_{ba}$ (red dashed curves) exhibit very different behaviors
except the point $k=\phi$ for given conditions. The ratio $%
I_{ab}/I_{ba} $ for scattering flows $I_{ab}$ and $I_{ba} $
is shown in Fig.~\ref{fig3}(d). In the regime for $k\neq \phi
$ ($0<\phi<\pi $), $I_{ab}/I_{ba}$ goes down with the increase of $J_{c}$
and $\gamma$. So the bandwidth of the region of nonreciprocity can be effectively broadened by increasing $J_{c}$ and $\gamma$ [satisfying Eq.~(\ref{Eq28})] simultaneously.

According to Eq.~(\ref{Eq27}), when $\Delta\neq 0$, i.e., the dissipative cavity
is not resonant with the cavities in the CRWs, we can still obtain perfect
nonreciprocity at the point $k=\phi$ ($0<\phi<\pi$) or $k=2\pi-\phi$ ($%
\pi<\phi<2\pi$) when the parameters satisfy the condition
\begin{equation}\label{Eq29}
J_{c}=\sqrt{ \frac{\Delta \xi}{\cos \phi } +2\xi ^{2}},
\end{equation}
\begin{equation}\label{Eq30}
\gamma =\xi ^{-1}J_{c}^{2}\sin \phi.
\end{equation}
To ensure that $J_{c}$ is real, the detuning $\Delta$ should be in the the
regime $\Delta>-2\xi\cos\phi$ ($0<\phi<\pi$) or $\Delta<-2\xi\cos\phi$ ($%
\pi<\phi<2\pi$). In Figs.~\ref{fig3}(e)-(g), we show the scattering flows $%
I_{ab}$ (black solid curves) and $I_{ba}$ (red dashed curves) as functions of
the wave number $k/\pi$ for different detunings $\Delta$
with phase $\phi=\pi/3$, where $J_{c}$ and $\gamma$ are chosen according to
Eqs.~(\ref{Eq29}) and (\ref{Eq30}). The scattering flows $I_{ab}$ and $I_{ba}$ exhibit very
different behaviors except the point $k=\phi$ for the different detunings $\Delta$.
The ratio $I_{ab}/I_{ba}$ of scattering flows $I_{ab}$ and $I_{ba}$ is shown in Fig.~\ref{fig3}(h). In the regime of $k\neq \phi$
($0<\phi<\pi$), $I_{ab}/I_{ba}$ also goes down (i.e., the bandwidth of the region of nonreciprocity becomes broader) with the increase of detuning $\Delta$ ($J_{c}$ and $\gamma$ increase accordingly).

\section{Single-photon circulator in T-shaped waveguide}

\subsection{Theoretical model and scattering matrix}

\begin{figure}[tbp]
\includegraphics[bb=0 272 590 641, width=8.5 cm, clip]{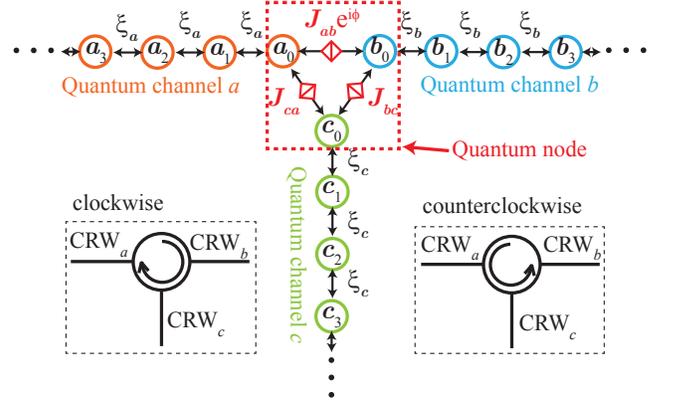}
\caption{(Color online) Schematic diagram of a T-shaped waveguide consisting
of three pairwise mutually coupling semi-infinite CRWs ($l_{j}$ for $%
l=a,b,c$ and $j\geq 0$).}
\label{fig4}
\end{figure}

Based on the single-photon nonreciprocity discussed in Sec.~II, we now study a three-port circulator for single photons in a T-shaped waveguide, which is constructed by replacing the dissipative cavity shown in Fig.~\ref{fig1} with a semi-infinite CRW (quantum channel $c$), as
schematically shown in Fig.~\ref{fig4}. The T-shaped waveguide can be
described by the Hamiltonian of Eq.~(\ref{Eq1}) with $H_{c}=\left( \omega _{c}-i\gamma \right) c_{0}^{\dag }c_{0}$ in Eq.~(\ref{Eq4}) replaced by
\begin{equation}
H_{c}=\omega _{c}\sum_{j=0}^{+\infty }c_{j}^{\dag }c_{j}-\xi
_{c}\sum_{j=0}^{+\infty }\left( c_{j}^{\dag }c_{j+1}+\mathrm{H.c.}\right),
\end{equation}%
where $c_{j}$ ($c_{j}^{\dag }$) is the bosonic annihilation (creation)
operator of the $j$th cavity in the CRW-$c$ with the same resonance frequency $%
\omega _{c}$ and the same coupling strength $\xi _{c}$ between two nearest-neighbor
cavities.

The stationary eigenstate of single-photon scattering in the T-shaped
waveguide is given by
\begin{equation}
\left\vert E\right\rangle =\sum_{l=a,b,c}\sum_{j=0}^{+\infty }u_{l}\left(
j\right) l_{j}^{\dag }\left\vert 0\right\rangle
\end{equation}%
where $\left\vert 0\right\rangle $ is the vacuum state of the T-shaped
waveguide, $l_{j}^{\dag }$ denotes the creation operator of the $j$th cavity
and $u_{l}\left( j\right) $ denotes the probability amplitude in the
state with a single photon in the $j$th cavity of the CRW-$l$. The
dispersion relation of the CRW-$c$ is given by~\cite{LZhouPRL13a}
\begin{equation}
E_{c}=\omega _{c}-2\xi _{c}\cos k_{c}, \quad 0<k_{c}<\pi,
\end{equation}%
where $E_{c}$ is the energy of the single-photon and $k_{c}$ is
the wave number of the single-photon in the CRW-$c$. Substituting the
stationary eigenstate and the Hamiltonian into the
eigenequation $H\left\vert E\right\rangle =E\left\vert E\right\rangle $, we
can obtain the coupled equations for the probability amplitudes in the
quantum node and quantum channels as in Eqs.~(\ref{Eq8})-(\ref{Eq11}) with Eq.(\ref{Eq10}) replaced by%
\begin{equation}
(\omega _{c}-E)u_{c}\left( 0\right) -\xi _{c}u_{c}\left( 1\right)
+J_{bc}u_{b}\left( 0\right) +J_{ca}u_{a}\left( 0\right) =0 ,
\end{equation}%
and the subscript $l=a,b$ in Eq.~(\ref{Eq11}) replaced by $l=a,b,c$.

\begin{widetext}
\begin{figure*}[tbp]
\includegraphics[bb=48 171 571 590, width=12.75 cm, clip]{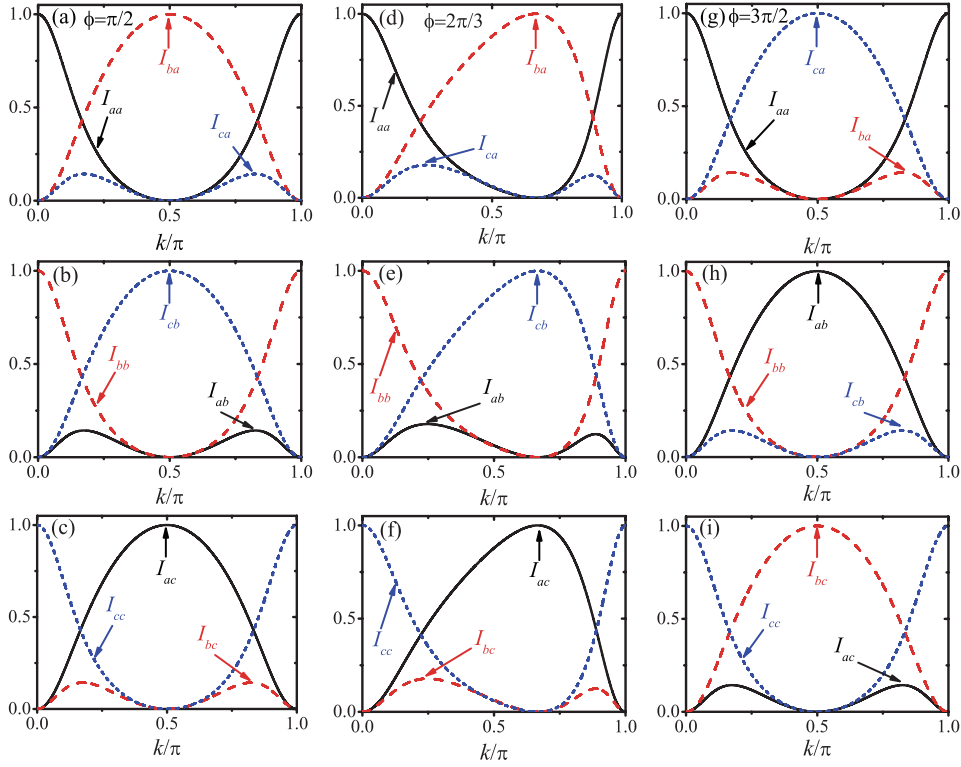}
\caption{(Color online) Scattering flows $I_{l^{\prime }l}$ ($l,l^{\prime }=a,b,c$) as functions of the wave number $k/\protect\pi$
for different phases: (a)-(c) $\phi=\pi/2$, (d)-(f) $\phi=2\pi/3$, and (g)-(i) $\phi=3\pi/2$. The other parameters are $\xi_{c}=J_{c}=J_{ab}=\xi$ and $k=k_{c}$.}
\label{fig5}
\end{figure*}
\end{widetext}

If a single photon with energy $E$ is incident from infinity side of CRW-$l$%
, the quantum node with the interactions among cavities $a_{0}$, $b_{0}$ and $c_{0}$ will result in photon sacttering between
different quantum channels. The general expressions for the probability
amplitudes in the three quantum channels ($l=a,b,c$) are given by ($j\geq 0$%
)
\begin{equation}
u_{l}\left( j\right) =e^{-ik_{l}j}+s_{ll}e^{ik_{l}j}, \label{Eq35}
\end{equation}
\begin{equation}
u_{l^{\prime }}\left( j\right) =s_{l^{\prime }l}e^{ik_{l^{\prime }}j}, \label{Eq36}
\end{equation}
\begin{equation}
u_{l^{\prime \prime }}\left( j\right) =s_{l^{\prime \prime
}l}e^{ik_{l^{\prime \prime }}j},\label{Eq37}
\end{equation}
where $s_{l^{\prime }l}$ ($s_{l^{\prime \prime }l}$) denotes the scattering
amplitude from the CRW-$l$ to the CRW-$l^{\prime }$ (CRW-$l^{\prime \prime }$%
). Substituting Eqs.~(\ref{Eq35})-(\ref{Eq37}) into the coupled equations for the probability
amplitudes then we obtain the scattering matrix as%
\begin{equation}
S=M^{-1}N,
\end{equation}%
where%
\begin{equation}
S=\left(
\begin{array}{ccc}
s_{aa} & s_{ab} & s_{ac} \\
s_{ba} & s_{bb} & s_{bc} \\
s_{ca} & s_{cb} & s_{cc}%
\end{array}%
\right),
\end{equation}%
\begin{equation}
M=\left(
\begin{array}{ccc}
\xi _{a}e^{-ik_{a}} & J_{ab}e^{i\phi} & J_{ca} \\
J_{ab}e^{-i\phi } & \xi _{b}e^{-ik_{b}} & J_{bc} \\
J_{ca} & J_{bc} & \xi _{c}e^{-ik_{c}}%
\end{array}%
\right),
\end{equation}%
\begin{equation}
N=\left(
\begin{array}{ccc}
-\xi _{a}e^{ik_{a}} & -J_{ab}e^{i\phi} & -J_{ca} \\
-J_{ab}e^{-i\phi} & -\xi _{b}e^{ik_{b}} & -J_{bc} \\
-J_{ca} & -J_{bc} & -\xi _{c}e^{ik_{c}}%
\end{array}%
\right).
\end{equation}

\subsection{Single-photon circulator}

\begin{figure}[tbp]
\includegraphics[bb=86 173 527 713, width=8.5 cm, clip]{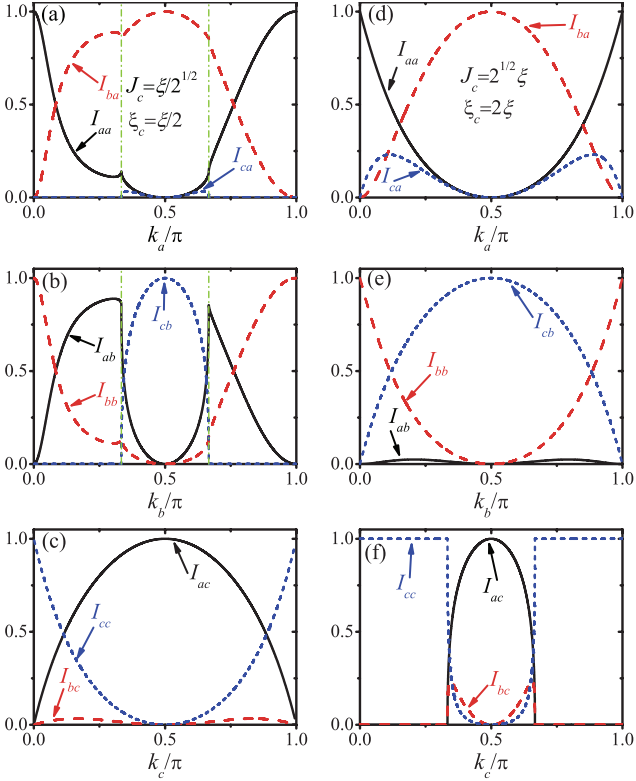}
\caption{(Color online) Scattering flows $I_{l^{\prime }l}$ ($l,l^{\prime
}=a,b,c$) as functions of the wave number $k_{l}/\protect\pi$ of a single photon incident from CRW-$l$ for (a)-(c) $%
\protect\xi_{c}=\protect\xi/2$ and $J_{c}=\protect\xi/\protect\sqrt{2}$,
(d)-(f) $\protect\xi_{c}=2\protect\xi$ and $J_{c}=\protect\sqrt{2}%
\protect\xi$. The other parameters are $J_{ab}=\protect\xi$ and $\protect\phi%
=\protect\pi/2$.}
\label{fig6}
\end{figure}

Before the numerical calculations of the scattering flows $I_{l^{\prime }l}$ ($l,l^{\prime }=a,b,c$), it is instructive to find the optimal conditions for observing perfect circulators.
A perfect circulator is obtained when we have $I_{ba}=I_{cb}=I_{ca}=1$ or $I_{ab}=I_{bc}=I_{ca}=1$
with all of the other scattering flows equal to zero.
For the sake of simplicity, we assume that all the coupled cavities in the T-shaped waveguide have the
same resonant frequency ($\omega _{a}=\omega _{b}=\omega _{c}$), the CRW-$a$
and CRW-$b$ have the same parameters (i.e. $\xi \equiv \xi _{a}=\xi _{b}$, $%
k\equiv k_{a}=k_{b}$) with the coupling strength $J_{ab}=\xi $, and they are
symmetrically coupled to CRW-$c$ (i.e. $J_{c}\equiv J_{bc}=J_{ca}$). On
these assumptions, the scattering amplitudes are obtained analytically as shown in Appendix~\ref{SPA-T}.
The optimal conditions for perfect circulator are summarized as follows.
If the coupling strengths $J_{c}$, $\xi _{c}$, and $\xi$ are equal, i.e.,
\begin{equation}
J_{c}=\xi _{c}=\xi,
\end{equation}
then the perfect circulator is obtained for
\begin{equation}
k=k_{c}=\left\{
\begin{array}{cc}
\phi  & 0<\phi <\pi  \\
2\pi -\phi  & \pi <\phi <2\pi
\end{array}%
\right..
\end{equation}%
However, if $J_{c}\neq \xi _{c}\neq \xi $, then the perfect circulator can only be obtained at
\begin{equation}
k=k_{c}=\frac{\pi }{2},
\end{equation}%
\begin{equation}
\phi =\frac{\pi }{2}\quad \mathrm{{or}\quad \frac{3\pi }{2}}
\end{equation}
with the coupling strengths $J_{c}$, $\xi _{c}$, and $\xi$ satisfying the following condition
\begin{equation}
J_{c}^{2}=\xi \xi _{c}.
\end{equation}%
This condition is consistent with the condition for single-photon nonreciprocity in the case of two CRWs with a dissipative cavity as given in Eq.~(\ref{Eq28}).

In Fig.~\ref{fig5}, the scattering flows $I_{o^{\prime }o}$ ($o,o^{\prime
}=a,b,c$) are plotted as functions of the wave number $k/\pi$ for different
phases with equal coupling strengths $\xi_{c}=J_{c}=\xi$. As shown in Figs.~\ref{fig5}%
(a)-(f), when $0<\phi<\pi$, we have $I_{ba}=I_{cb}=I_{ac}=1$ with the other
scattering flows which are equal to zero for the wave number $k=\phi$. As shown in Figs.~\ref%
{fig5}(g)-(i), when $\pi<\phi<2\pi$, we have $I_{ab}=I_{bc}=I_{ca}=1$ with
the other scattering flows which are equal to zero for the wave number $k=2\pi-\phi$. In other words,
when $0<\phi<\pi$, the signal is transferred from one CRW to another
clockwise ($a\rightarrow b \rightarrow c \rightarrow a$) for $%
k=\phi $. On the contrary, when $\pi<\phi<2\pi$, the signal is transferred
from one CRW to another counterclockwise ($a\rightarrow c \rightarrow b
\rightarrow a$) for $k=2\pi-\phi$.

When $\xi_{c}\neq\xi$, the band widths of CRW-$a$ and CRW-$b$ are different from
the band width of CRW-$c$, and nonreciprocity ($I_{l^{\prime }l}\neq
I_{ll^{\prime }}$) can only be obtained in the overlap band regime among the
three CRWs.  In this case, we can have
perfect circulator ($I_{ba}=I_{cb}=I_{ac}=1$ with the other zero scattering flows) when $k=\pi/2$ for $\phi=\pi/2$ or $3\pi/2$.
In Fig.~\ref{fig6}, the scattering flows $I_{l^{\prime }l}$ ($l,l^{\prime
}=a,b,c$) are plotted as functions of the wave number $k/\pi$ with different coupling strengths
$\xi_{c}\neq J_{c} \neq \xi$.
As shown in Figs.~\ref{fig6}(a)-(c), when $\xi_{c}=\xi/2$ and $%
J_{c}=\xi/\sqrt{2}$, the scattering flows with $I_{l^{\prime }l}\neq I_{ll^{\prime }}$ appear in the
regime $\pi/3<k_{a}=k_{b}<2\pi/3$ and $0<k_{c}<\pi$, and we have $I_{ba}=
I_{ab}$ in the regimes of $0<k_{a}=k_{b}<\pi/3$ and $2\pi/3<k_{a}=k_{b}<\pi$%
. As shown in Figs.~\ref{fig6}(d)-(f), when $\xi_{c}=2\xi$ and $J_{c}=\sqrt{2}%
\xi$, we have nonreciprocity in the regime of $0<k_{a}=k_{b}<\pi$ and $%
\pi/3<k_{c}<2\pi/3$, the single photon incident from the infinity side of CRW-$c$ will be reflected totally ($%
I_{cc}=1$) in the regimes of $0<k_{c}<\pi/3$ and $2\pi/3<k_{c}<\pi$.

\section{Conclusions}

In summary, we have shown that a single photon can nonreciprocally
transport between two coupled
one-dimensional semi-infinite CRWs, in which
both end sides are coupled to a dissipative cavity. Moreover, we have shown that a T-shaped
waveguide consisting of three pairwise mutually coupled semi-infinite
CRWs can be used as a single-photon circulator. The optimal conditions for
the single-photon nonreciprocity and the single-photon circulator are given
analytically. The CRWs connected by quantum node with broken
time-reversal symmetry will open up a kind of quantum devices with versatile applications in the quantum
networks.

Finally, let us provide some remarks on the experimental feasibility of our
proposal. In our model, the key element is the quantum node of the
three coupled cavity modes with relative phase, which have already been realized
experimentally with superconducting Josephson junctions~\cite{LecocqarX16}
and optomechanical (electromechanical) interactions~\cite{BernierarX16}
recently. Moreover, the coupled cavity modes also can be replaced by the
coupling qubits~\cite{Roushan16a}, because they are equivalent at the
single-photon level. The quantum channels (CRWs) can be realized by using
coupled superconducting transmission line resonators~\cite{WallraffNat04},
superconducting quantum interference device arrays~\cite{CastellanosAPL07},
or defect resonators in two-dimensional photonic crystals~\cite{NotomiNPT08}%
. Thus, if the quantum nodes can be connected by quantum channels,
then the single-photon nonreciprocity and circulator may be demonstrated
in the proposed system, and this may be an important step forward in the
realization of the quantum networks.

\vskip 2pc \leftline{\bf Note added} After finishing this work, a preprint studying nonreciprocal propagation in linear time-invariant waveguides appears on arXiv~\cite{PetersonarXiv}.

\vskip 2pc \leftline{\bf Acknowledgement}

We thank Lan Zhou and Zhi-Hai Wang for fruitful discussions.
X.-W.X. is supported by the National Natural Science Foundation of China
(NSFC) under Grants No.11604096 and the Startup Foundation for Doctors of
East China Jiaotong University under Grant No. 26541059. A.-X.C. is
supported by NSFC under Grants No. 11365009.
Y.L. is supported by NSFC under Grants No. 11421063.
Y.L. and Y.-X.L. are supported by the National Basic Research Program of China (973 Program) under
Grants No. 2014CB921400.
Y.-X.L. is also supported by NSFC under Grant Nos.
61328502 and 61025022, the Tsinghua University Initiative Scientific
Research Program, and the Tsinghua National Laboratory for Information
Science and Technology (TNList) Cross-discipline Foundation.

\appendix

\section{Scattering probability amplitudes for T-shaped waveguide}\label{SPA-T}

On the assumptions, $\omega _{a}=\omega _{b}=\omega _{c}$, $\xi \equiv \xi _{a}=\xi _{b}=J_{ab}$, $%
k\equiv k_{a}=k_{b}$, $J_{c}\equiv J_{bc}=J_{ca}$, the scattering amplitudes can be given analytically
as follows,
\begin{equation}
s_{aa}=-\xi J_{c}^{2}\left( e^{i\phi }+e^{-i\phi }-e^{-ik}-e^{ik}\right)
D^{-1},
\end{equation}%
\begin{equation}
s_{ba}=\xi e^{-ik_{c}}\left( \xi \xi _{c}e^{-i\phi
}-J_{c}^{2}e^{ik_{c}}\right) \left( e^{ik_{a}}-e^{-ik_{a}}\right) D^{-1},
\end{equation}%
\begin{equation}
s_{ca}=\xi ^{2}J_{c}\left( e^{-ik}-e^{ik}\right) \left( e^{-i\phi
}-e^{-ik}\right) D^{-1},
\end{equation}%
\begin{equation}
s_{ab}=\xi e^{-ik_{c}}\left( \xi \xi _{c}e^{i\phi
}-J_{c}^{2}e^{ik_{c}}\right) \left( e^{ik}-e^{-ik}\right) D^{-1},
\end{equation}%
\begin{equation}
s_{bb}=-\xi J_{c}^{2}\left( e^{i\phi }+e^{-i\phi }-e^{ik}-e^{-ik}\right)
D^{-1},
\end{equation}%
\begin{equation}
s_{cb}=\xi ^{2}J_{c}\left( e^{-ik}-e^{ik}\right) \left( e^{i\phi
}-e^{-ik}\right) D^{-1},
\end{equation}%
\begin{equation}
s_{ac}=\xi J_{c}\xi _{c}\left( e^{-ik_{c}}-e^{ik_{c}}\right) \left( e^{i\phi
}-e^{-ik}\right) D^{-1},
\end{equation}%
\begin{equation}
s_{bc}=\xi J_{c}\xi _{c}\left( e^{-ik_{c}}-e^{ik_{c}}\right) \left(
e^{-i\phi }-e^{-ik}\right) D^{-1},
\end{equation}%
\begin{equation}
s_{cc}=\left[ \xi J_{c}^{2}\left( 2e^{-ik}-e^{i\phi }-e^{-i\phi }\right)
+\xi ^{2}\xi _{c}e^{ik_{c}}\left( 1-e^{-i2k}\right) \right] D^{-1},
\end{equation}%
\begin{equation}
D=\xi J_{c}^{2}\left( e^{i\phi }+e^{-i\phi }-2e^{-ik}\right) +\xi ^{2}\xi
_{c}e^{-ik_{c}}\left( e^{-i2k}-1\right).
\end{equation}%

\bibliographystyle{apsrev}
\bibliography{ref}

\begin{thebibliography}{99}
\bibitem{KimbleNat08a} H. J. Kimble, The quantum internet, Nature (London)~%
\textbf{453}, 1023 (2008).

\bibitem{WallraffNat04} A. Wallraff, D. I. Schuster, A. Blais, L. Frunzio,
R. S. Huang, J. Majer, S. Kumar, S. M. Girvin, and R. J. Schoelkopf, Strong
coupling of a single photon to a superconducting qubit using circuit quantum
electrodynamics, Nature (London)~\textbf{431}, 162 (2004).

\bibitem{CastellanosAPL07} M. A. Castellanos-Beltran and K. W. Lehnert,
Widely tunable parametric amplifier based on a superconducting quantum
interference device array resonator, Appl. Phys. Lett.~\textbf{91}, 083509
(2007)

\bibitem{NotomiNPT08} M. Notomi, E. Kuramochi, and T. Tanabe, Large-scale
arrays of ultrahigh-Q coupled nanocavities, Nat. Photon.~\textbf{2}, 741
(2008).

\bibitem{AbdumalikovPRL10a} A. A. Abdumalikov Jr., O. Astafiev, A. M.
Zagoskin, Y. A. Pashkin, Y. Nakamura, and J. S. Tsai, Electromagnetically
Induced Transparency on a Single Artificial Atom, Phys. Rev. Lett.~\textbf{%
104}, 193601 (2010).

\bibitem{LZhouPRL08a} L. Zhou, Z. R. Gong, Y. X. Liu, C. P. Sun, and F.
Nori, Controllable Scattering of a Single Photon inside a One-Dimensional
Resonator Waveguide, Phys. Rev. Lett.~\textbf{101}, 100501 (2008).

\bibitem{LZhouPRA08a} L. Zhou, H. Dong, Y. X. Liu, C. P. Sun, and F. Nori,
Quantum supercavity with atomic mirrors, Phys. Rev. A~\textbf{78}, 063827
(2008).

\bibitem{JQLiaoPRA09a} J. Q. Liao, J. F. Huang, Y. X. Liu, L. M. Kuang, and
C. P. Sun, Quantum switch for single-photon transport in a coupled
superconducting transmission-line-resonator array, Phys. Rev. A~\textbf{80},
014301 (2009).

\bibitem{LZhouPRA09a} L. Zhou, S. Yang, Y. X. Liu, C. P. Sun, and F. Nori,
Quantum Zeno switch for single-photon coherent transport, Phys. Rev. A~%
\textbf{80}, 062109 (2009).

\bibitem{LongoPRL10a} P. Longo, P. Schmitteckert, and K. Busch, Few-Photon Transport in Low-Dimensional Systems: Interaction-Induced Radiation Trapping, Phys. Rev. Lett.~\textbf{104}, 023602 (2010).

\bibitem{JQLiaoPRA10a} J. Q. Liao, Z. R. Gong, L. Zhou, Y. X. Liu, C. P.
Sun, and F. Nori, Controlling the transport of single photons by tuning the
frequency of either one or two cavities in an array of coupled cavities,
Phys. Rev. A~\textbf{81}, 042304 (2010).

\bibitem{ZHWangPRA12a} Z. H. Wang, Y. Li, D. L. Zhou, C. P. Sun, and P.
Zhang, Single-photon scattering on a strongly dressed atom, Phys. Rev. A~%
\textbf{86}, 023824 (2012).

\bibitem{LZhouPRA12a} L. Zhou, Y. Chang, H. Dong, L. M. Kuang, and C. P.
Sun, Inherent Mach-Zehnder interference with ``which-way'\ detection for
single-particle scattering in one dimension, Phys. Rev. A~\textbf{85},
013806 (2012).

\bibitem{MTChengPRa12a} M. T. Cheng, X. S. Ma, M. T. Ding, Y. Q. Luo, and G.
X. Zhao, Single-photon transport in one-dimensional coupled-resonator
waveguide with local and nonlocal coupling to a nanocavity containing a
two-level system, Phys. Rev. A~\textbf{85}, 053840 (2012).

\bibitem{ZRGongPRA08a} Z. R. Gong, H. Ian, L. Zhou, and C. P. Sun,
Controlling quasibound states in a one-dimensional continuum through an
electromagnetically-induced-transparency mechanism, Phys. Rev. A~\textbf{78}%
, 053806 (2008).

\bibitem{LZhouPRL13a} L. Zhou, L. P. Yang, Y. Li, and C. P. Sun, Quantum
Routing of Single Photons with a Cyclic Three-Level System, Phys. Rev. Lett.~%
\textbf{111}, 103604 (2013).

\bibitem{JLuPRA14a} J. Lu, L. Zhou, L. M. Kuang, and F. Nori, Single-photon
router: Coherent control of multichannel scattering for single photons with
quantum interferences, Phys. Rev. A~\textbf{89}, 013805 (2014).

\bibitem{JLuOE15a} J. Lu, Z. H. Wang, and L. Zhou, T-shaped single-photon
router, Opt. Express~\textbf{23}, 22955 (2015).

\bibitem{ZHWangPRA14a} Z. H. Wang, L. Zhou, Y. Li, and C. P. Sun,
Controllable single-photon frequency converter via a one-dimensional
waveguide, Phys. Rev. A~\textbf{89}, 053813 (2014).

\bibitem{JalasNPT13} D. Jalas, A. Petrov, M. Eich, W. Freude, S. Fan, Z. Yu,
R. Baets, M. Popovi\'{c}, A. Melloni, J. D. Joannopoulos, M. Vanwolleghem,
C. R. Doerr, and H. Renner, What is - and what is not - an optical isolator,
Nat. Photon.~\textbf{7}, 579 (2013).

\bibitem{FujitaAPL00} J. Fujita, M. Levy, R. M. Osgood, L.Wilkens, and H. D%
\"{o}tsch, Waveguide optical isolator based on Mach-Zehnder interferometer,
Appl. Phys. Lett.~\textbf{76}, 2158 (2000).

\bibitem{EspinolaOL04} R. L. Espinola, T. Izuhara, M. C. Tsai, R. M. Osgood
Jr., H. D\"{o}tsch, Magneto-optical nonreciprocal phase shift in
garnet/silicon-on-insulator waveguides, Opt. Lett.~\textbf{29}, 941 (2004).

\bibitem{ZamanAPL07} T. R. Zaman, X. Guo, R. J. Ram, Faraday rotation in an
InP waveguide, Appl. Phys. Lett.~\textbf{90}, 023514 (2007).

\bibitem{HaldanePRL08} F. D. M. Haldane and S. Raghu, Possible Realization
of Directional Optical Waveguides in Photonic Crystals with Broken
Time-Reversal Symmetry, Phys. Rev. Lett.~\textbf{100}, 013904 (2008).

\bibitem{ShojiAPL08} Y. Shoji, T. Mizumoto, H. Yokoi, I. Hsieh, and R. M.
Osgood Jr., Magneto-optical isolator with silicon waveguides fabricated by
direct bonding, Appl. Phys. Lett.~\textbf{92}, 071117 (2008).

\bibitem{ZWangNat09} Z. Wang, Y. Chong, J. D. Joannopoulos, and M. Solja\v{c}%
i\'{c}, Observation of unidirectional backscattering-immune topological
electromagnetic states, Nature (London)~\textbf{461}, 772 (2009).

\bibitem{HadadPRL10} Y. Hadad and B. Z. Steinberg, Magnetized Spiral Chains
of Plasmonic Ellipsoids for One-Way Optical Waveguides, Phys. Rev. Lett.~%
\textbf{105}, 233904 (2010).

\bibitem{KhanikaevPRL10} A. B. Khanikaev, S. H. Mousavi, G. Shvets, and Y.
S. Kivshar, One-Way Extraordinary Optical Transmission and Nonreciprocal
Spoof Plasmons, Phys. Rev. Lett.~\textbf{105}, 126804 (2010).

\bibitem{LBiNPo11} L. Bi, J. Hu, P. Jiang, D. H. Kim, G. F. Dionne, L. C.
Kimerling, and C. A. Ross, On-chip optical isolation in monolithically
integrated non-reciprocal optical resonators, Nat. Photon.~\textbf{5}, 758
(2011).

\bibitem{ShojiOE12} Y. Shoji, M. Ito, Y. Shirato, and T. Mizumoto, MZI
optical isolator with Si-wire waveguides by surface-activated direct
bonding, Opt. Express~\textbf{20}, 18440 (2012).

\bibitem{ZWangSR15a} Z. Wang, L. Shi, Y. Liu, X. Xu, and X. Zhang, Optical
Nonreciprocity in Asymmetric Optomechanical Couplers. Sci. Rep.~\textbf{5},
8657 (2015).

\bibitem{YShiNPt15a} Y. Shi, Z. Yu, and S. Fan, Limitations of nonlinear
optical isolators due to dynamic reciprocity, Nat. Photon~\textbf{9}, 388
(2015).

\bibitem{XGuoPRL16} X. Guo, C. L. Zou, H. Jung, and H. X. Tang, On-Chip
Strong Coupling and Efficient Frequency Conversion between Telecom and
Visible Optical Modes, Phys. Rev. Lett.~\textbf{117}, 123902 (2016).

\bibitem{QTCaoPRL17} Q. T. Cao, H. Wang, C. H. Dong, H. Jing, R. S. Liu, X.
Chen, L. Ge, Q. Gong, and Y. F. Xiao, Experimental Demonstration of
Spontaneous Chirality in a Nonlinear Microresonator, Phys. Rev. Lett.~%
\textbf{118}, 033901 (2017).

\bibitem{DelBinoarX16a} L. Del Bino, J. M. Silver, S. L. Stebbings, and P.
Del'Haye, Symmetry Breaking of Counter-Propagating Light in a Nonlinear
Resonator, arXiv:1607.01194 [physics.optics].

\bibitem{GalloAPL01} K. Gallo, G. Assanto, K. R. Parameswaran, and M. M.
Fejer, All-optical diode in a periodically poled lithium niobate waveguide,
Appl. Phys. Lett.~\textbf{79}, 314 (2001).

\bibitem{MingaleevJOSAB02} S. F. Mingaleev, Y. S. Kivshar, Nonlinear
transmission and light localization in photonic-crystal waveguides, J. Opt.
Soc. Am. B~\textbf{19}, 2241 (2002).

\bibitem{SoljacicOL03} M. Solja\v{c}i\'{c}, C. Luo, J. D. Joannopoulos, S.
Fan, Nonlinear photonic crystal microdevices for optical integration, Opt.
Lett.~\textbf{28}, 637 (2003).

\bibitem{RostamiOLT07} A. Rostami, Piecewise linear integrated optical
device as an optical isolator using two-port nonlinear ring resonators, Opt.
Laser Technol.~\textbf{39}, 1059 (2007).

\bibitem{AlberucciOL08} A. Alberucci and G. Assanto, All-optical isolation
by directional coupling, Opt. Lett.~\textbf{33}, 1641 (2008).

\bibitem{LFanSci12} L. Fan, J. Wang, L. T. Varghese, H. Shen, B. Niu, Y.
Xuan, A. M. Weiner, and M. Qi, An All-Silicon Passive Optical Diode, Science~%
\textbf{335}, 447 (2012).

\bibitem{LFanOL13} L. Fan, L. T. Varghese, J. Wang, Y. Xuan, A. M. Weiner,
and M. Qi, Silicon optical diode with 40 dB nonreciprocal transmission, Opt.
Lett.~\textbf{38}, 1259 (2013).

\bibitem{AnandNL13} B. Anand, R. Podila, K. Lingam, S. R. Krishnan, S. S. S.
Sai, R. Philip, and A. M. Rao, Optical Diode Action from Axially Asymmetric
Nonlinearity in an All-Carbon Solid-State Device, Nano Lett.~\textbf{13},
5771 (2013).

\bibitem{BiancalanaJAP08} F. Biancalana, All-optical diode action with
quasiperiodic photonic crystals, J. Appl. Phys.~\textbf{104}, 093113 (2008).

\bibitem{MiroshnichenkoAPL10} A. E. Miroshnichenko, E. Brasselet, and Y. S.
Kivshar, Reversible optical nonreciprocity in periodic structures with
liquid crystals, Appl. Phys. Lett.~\textbf{96}, 063302 (2010).

\bibitem{CWangOE11} C. Wang, C. Zhou, and Z. Li, On-chip optical diode based
on silicon photonic crystal heterojunctions, Opt. Express~\textbf{19}, 26948
(2011).

\bibitem{CWangSR12} C. Wang, X. Zhong, and Z. Li, Linear and passive silicon
optical isolator, Sci. Rep.~\textbf{2}, 674 (2012).

\bibitem{KXiaOE13} K. Xia, M. Alamri, and M. S. Zubairy, Ultrabroadband
nonreciprocal transverse energy flow of light in linear passive photonic
circuits, Opt. Express~\textbf{21}, 25619 (2013).

\bibitem{LenferinkOE14} E. J. Lenferink, G. Wei, and N. P. Stern, Coherent
optical non-reciprocity in axisymmetric resonators, Opt. Express~\textbf{22}%
, 16099 (2014).

\bibitem{YYuarX14} Y. Yu, Y. Chen, H. Hu, W. Xue, K. Yvind, and J. Mork,
Nonreciprocal transmission in a nonlinear photonic-crystal Fano structure
with broken symmetry, Laser Photonics Rev.~\textbf{9}, 241 (2015).

\bibitem{YuNP09} Z. F. Yu and S. H. Fan, Complete optical isolation created
by indirect interband photonic transitions, Nat. Photon.~\textbf{3}, 91
(2009).

\bibitem{KFangNPo12} K. Fang, Z. Yu, and S. Fan, Realizing effective
magnetic field for photons by controlling the phase of dynamic modulation,
Nat. Photon.~\textbf{6}, 782 (2012).

\bibitem{ELiNC14} E. Li, B. J. Eggleton, K. Fang, and S. Fan, Photonic
Aharonov-Bohm effect in photon-phonon interactions, Nat. Commun.~\textbf{5},
3225 (2014).

\bibitem{DoerrOL11} C. R. Doerr, N. Dupuis, and L. Zhang, Optical isolator
using two tandem phase modulators, Opt. Lett.~\textbf{36}, 4293 (2011).

\bibitem{DoerrOE14} C. R. Doerr, L. Chen, and D. Vermeulen, Silicon
photonics broadband modulation-based isolator, Opt. Express~\textbf{22},
4493 (2014).

\bibitem{LiraPRL12} H. Lira, Z. F. Yu, S. H. Fan, and M. Lipson,
Electrically Driven Nonreciprocity Induced by Interband Photonic Transition
on a Silicon Chip, Phys. Rev. Lett.~\textbf{109}, 033901 (2012).

\bibitem{TzuangNPt14} L. D. Tzuang, K. Fang, P. Nussenzveig, S. Fan, and M.
Lipson, Non-reciprocal phase shift induced by an effective magnetic flux for
light, Nat. Photon.~\textbf{8}, 701 (2014).

\bibitem{MunozPRL14} M. Castellanos Munoz, A. Y. Petrov, L. O'Faolain, J.
Li, T. F. Krauss, and M. Eich, Optically Induced Indirect Photonic
Transitions in a Slow Light Photonic Crystal Waveguide, Phys. Rev. Lett.~%
\textbf{112}, 053904 (2014).

\bibitem{YYangOE14} Y. Yang, C. Galland, Y. Liu, K. Tan, R. Ding, Q. Li, K.
Bergman, T. Baehr-Jones, and M. Hochberg, Experimental demonstration of
broadband Lorentz non-reciprocity in an integrable photonic architecture
based on Mach-Zehnder modulators, Opt. Express~\textbf{22}, 17409 (2014).

\bibitem{MSKangNP11} M. S. Kang, A. Butsch, and P. S. J. Russell,
Reconfigurable light-driven opto-acoustic isolators in photonic crystal
fibre, Nat. Photonics~\textbf{5}, 549 (2011).

\bibitem{EuterNP10} C. E\"{u}ter, K. G. Makris, R. EI-Ganainy, D. N.
Christodoulides, M. Segev, and D. Kip, Observation of parity-time symmetry
in optics, Nat. Phys.~\textbf{6}, 192 (2010).

\bibitem{LFengSci11} L. Feng, M. Ayache, J. Q. Huang, Y. L. Xu, M. H. Lu, Y.
F. Chen, Y. Fainman, and A. Scherer, Nonreciprocal Light Propagation in a
Silicon Photonic Circuit, Science~\textbf{333}, 729 (2011).

\bibitem{BPengNP14} B. Peng, S. K. \"{O}zdemir, F. Lei, F. Monifi, M.
Gianfreda, G. L. Long, S. H. Fan, F. Nori, C. M. Bender, and L. Yang,
Parity-time-symmetric whispering-gallery microcavities, Nat. Phys.~\textbf{10%
}, 394 (2014).

\bibitem{WangPRL13} D. W. Wang, H. T. Zhou, M. J. Guo, J. X. Zhang, J.
Evers, and S. Y. Zhu, Optical Diode Made from a Moving Photonic Crystal,
Phys. Rev. Lett.~\textbf{110}, 093901 (2013).

\bibitem{WangOE10} Q. Wang, F. Xu, Z. Y. Yu, X. S. Qian, X. K. Hu, Y. Q. Lu,
and H. T. Wang, A bidirectional tunable optical diode based on periodically
poled LiNbO3, Opt. Express~\textbf{18}, 7340 (2010).

\bibitem{RamezaniPRA10} H. Ramezani, T. Kottos, R. El-Ganainy, and D. N.
Christodoulides, Unidirectional nonlinear $\mathcal{PT} $-symmetric optical
structures, Phys. Rev. A~\textbf{82}, 043803 (2010).

\bibitem{KFangPRL12} K. Fang, Z. Yu, and S. Fan, Photonic Aharonov-Bohm
Effect Based on Dynamic Modulation, Phys. Rev. Lett.~\textbf{108}, 153901
(2012).

\bibitem{HorsleyPRL13} S. A. R. Horsley, J. H. Wu, M. Artoni, and G. C. La
Rocca, Optical Nonreciprocity of Cold Atom Bragg Mirrors in Motion, Phys.
Rev. Lett.~\textbf{110}, 223602 (2013).

\bibitem{JHWuPRL14} J. H. Wu, M. Artoni, and G. C. La Rocca, Non-Hermitian
Degeneracies and Unidirectional Reflectionless Atomic Lattices, Phys. Rev.
Lett.~\textbf{113}, 123004 (2014).

\bibitem{ManipatruniPRL09} S. Manipatruni, J. T. Robinson, and M. Lipson,
Optical Nonreciprocity in Optomechanical Structures, Phys. Rev. Lett.~%
\textbf{102}, 213903 (2009).

\bibitem{HafeziOE12} M. Hafezi and P. Rabl, Optomechanically induced
non-reciprocity in microring resonators, Opt. Express~\textbf{20}, 7672
(2012).

\bibitem{KimNPy15} J. Kim, M. C. Kuzyk, K. Han, H. Wang, and G. Bahl,
Non-reciprocal Brillouin scattering induced transparency, Nat. Phys.~\textbf{%
11}, 275 (2015).

\bibitem{CHDongNC15} C. H. Dong, Z. Shen, C. L. Zou, Y. L. Zhang, W. Fu, and
G. C. Guo, Brillouin-scattering-induced transparency and non-reciprocal
light storage, Nature Commun.~\textbf{6}, 6193 (2015).

\bibitem{EstepNPy14a} N. A. Estep, D. L. Sounas, J. Soric and A. Al\`{u},
Magnetic-free non-reciprocity and isolation based on parametrically
modulated coupled-resonator loops, Nat. Phys.~\textbf{10}, 923 (2014).

\bibitem{ZShenNPt16a} Z. Shen, Y. L. Zhang, Y. Chen, C. L. Zou, Y. F. Xiao,
X. B. Zou, F. W. Sun, G. C. Guo, and C. H. Dong, Experimental realization of
optomechanically induced non-reciprocity, Nature Photonics~\textbf{10}, 657
(2016).

\bibitem{FangKArx15} K. Fang, M. H. Matheny, X. Luan, and O. Painter,
Optical transduction and routing of microwave phonons in
cavity-optomechanical circuits, Nat. Photon.~\textbf{10}, 489 (2016).

\bibitem{KochPRA10} J. Koch, A. A. Houck, K. L. Hur, and S. M. Girvin,
Time-reversal-symmetry breaking in circuit-QED-based photon lattices, Phys.
Rev. A~\textbf{82}, 043811 (2010).

\bibitem{HabrakenNJP12} S. J. M. Habraken, K. Stannigel, M. D. Lukin, P.
Zoller, and P Rabl, Continuous mode cooling and phonon routers for phononic
quantum networks, New J. Phys.~\textbf{14}, 115004 (2012).

\bibitem{RanzaniNJP14a} L. Ranzani and J. Aumentado, A geometric description
of nonreciprocity in coupled two-mode systems, New J. Phys.~\textbf{16}, 103027
(2014).

\bibitem{RanzaniNJP15a} L. Ranzani and J. Aumentado, Graph-Based Analysis of
Nonreciprocity in Coupled-Mode Systems, New J. Phys.~\textbf{17}, 023024 (2015).

\bibitem{YPWangSR15a} Y. P. Wang, W. Wang, Z. Y. Xue, W. L. Yang, Y. Hu, and
Y. Wu, Realizing and characterizing chiral photon flow in a circuit quantum
electrodynamics necklace, Sci. Rep.~\textbf{5}, 8352 (2015).

\bibitem{SliwaPRX15} K. M. Sliwa, M. Hatridge, A. Narla, S. Shankar, L.
Frunzio, R. J. Schoelkopf, and M. H. Devoret, Reconfigurable Josephson
Circulator/Directional Amplifier, Phys. Rev. X~\textbf{5}, 041020 (2015).

\bibitem{SchmidtOpt15} M. Schmidt, S. Kessler, V. Peano, O. Painter, and F.
Marquardt, Optomechanical creation of magnetic fields for photons on a
lattice, Optica~\textbf{2}, 635 (2015).

\bibitem{MetelmannPRX15} A. Metelmann and A. A. Clerk, Nonreciprocal Photon
Transmission and Amplification via Reservoir Engineering, Phys. Rev. X~%
\textbf{5}, 021025 (2015).

\bibitem{XuXWPRA15} X. W. Xu and Y. Li, Optical nonreciprocity and
optomechanical circulator in three-mode optomechanical systems, Phys. Rev. A~%
\textbf{91}, 053854 (2015).

\bibitem{XWXuPRA16a} X. W. Xu, Y. Li, A. X. Chen, and Y. X. Liu,
Nonreciprocal conversion between microwave and optical photons in
electro-optomechanical systems, Phys. Rev. A~\textbf{93}, 023827 (2016).

\bibitem{MetelmannarX16a} A. Metelmann and A. A. Clerk, Non-reciprocal
quantum interactions and devices via autonomous feed-forward,
arXiv:1610.06621 [quant-ph].

\bibitem{LTianarx16a} L. Tian and Z. Li, Nonreciprocal State Conversion
between Microwave and Optical Photons, arXiv:1610.09556 [quant-ph].

\bibitem{MiriarX16a} M.-A. Miri, F. Ruesink, E. Verhagen, and A. Al\`{u},
Fundamentals of optical non-reciprocity based on optomechanical coupling,
arXiv:1612.07375 [physics.optics].

\bibitem{XXuarX17a} X. Xu and J. M. Taylor, Optomechanically-induced chiral
transport of phonons in one dimension, arXiv:1701.02699 [quant-ph].

\bibitem{RuesinkNC16a} F. Ruesink, M.-A. Miri, A. Al\`{u}, and E. Verhagen,
Nonreciprocity and magnetic-free isolation based on optomechanical
interactions. Nat. Commun.~\textbf{7}, 13662 (2016).

\bibitem{KFangNPy17a} K. Fang, J. Luo, A. Metelmann, M. H. Matheny, F.
Marquardt, A. A. Clerk, O. Painter, Generalized non-reciprocity in an
optomechanical circuit via synthetic magnetism and reservoir engineering,
Nature Physics (2017) (Published online) doi:10.1038/nphys4009.

\bibitem{BernierarX16a} N. R. Bernier, L. D. T\'{o}th, A. Koottandavida, A.
Nunnenkamp, A. K. Feofanov, T. J. Kippenberg, Nonreciprocal reconfigurable
microwave optomechanical circuit, arXiv:1612.08223 [quant-ph].

\bibitem{Gardiner} C. W. Gardiner and M. J. Collett, Input and output in damped quantum systems: Quantum stochastic differential equations and the master equation, Phys. Rev. A~\textbf{31}, 3761 (1985).

\bibitem{LecocqarX16} F. Lecocq, L. Ranzani, G. A. Peterson, K. Cicak, R. W.
Simmonds, J. D. Teufel, and J. Aumentado, Nonreciprocal microwave signal
processing with a Field-Programmable Josephson Amplifier, arXiv:1612.01438
[quant-ph].

\bibitem{BernierarX16} N. R. Bernier, L. D. T\'{o}th, A. Koottandavida, A.
Nunnenkamp, A. K. Feofanov, and T. J. Kippenberg, Nonreciprocal
reconfigurable microwave optomechanical circuit, arXiv:1612.08223 [quant-ph].

\bibitem{Roushan16a} P. Roushan, C. Neill, A. Megrant, Y. Chen, R. Babbush,
R. Barends, B. Campbell, Z. Chen, B. Chiaro, A. Dunsworth, A. Fowler, E.
Jeffrey, J. Kelly, E. Lucero, J. Mutus, P. J. J. O'Malley, M. Neeley, C.
Quintana, D. Sank, A. Vainsencher, J. Wenner, T. White, E. Kapit, H. Neven,
and J. Martinis, Chiral groundstate currents of interacting photons in a
synthetic magnetic field, Nature Physics (2016) (Published online).

\bibitem{PetersonarXiv} C. W. Peterson, S. Kim, J. T. Bernhard, and G. Bahl, Reconfigurable arbitrary nonreciprocal transfer functions through nonreciprocal coupling, arXiv:1702.06476 [physics.optics].


\end{thebibliography}

\end{document}